\documentclass[aps,pre,twocolumn,showpacs,superscriptaddress,groupedaddress]{revtex4}
\newcommand{\bea}{\begin{eqnarray}}
\newcommand{\eea}{\end{eqnarray}}
\newcommand{\beq}{\begin{equation}}
\newcommand{\eeq}{\end{equation}}
\newcommand{\ben}{\begin{enumerate}}
\newcommand{\een}{\end{enumerate}}
\newcommand{\bet}{\begin{split}}
\newcommand{\eet}{\end{split}}

\newcommand{\av}[1]{\langle{#1}\rangle}

\renewcommand{\>}{\rangle}
\newcommand{\comm}{\color{black}}

\usepackage{hyperref}
\usepackage{graphicx,color}
\usepackage{bm}
\usepackage{amsmath}
\usepackage{amssymb}
\usepackage{amsfonts}

\begin{document}
\title{The Effect of Phenotypic Selection on Stochastic Gene Expression}

\author{Thierry Mora}

\address{Laboratoire de Physique Statistique, CNRS,    Universit\'e P. et M. Curie,    \'Ecole Normale Sup\'erieure,  Paris, France.}

%\affiliation{Laboratoire de Physique Statistique, CNRS,    Universit\'e P. et M. Curie,    \'Ecole Normale Sup\'erieure,  Paris, France}

\author{Aleksandra M. Walczak}

\address{Laboratoire de Physique Th\'eorique, CNRS,    Universit\'e P. et M. Curie,    \'Ecole Normale Sup\'erieure,  Paris, France.}

%\affiliation{Laboratoire de Physique Th\'eorique, CNRS,    Universit\'e P. et M. Curie,    \'Ecole Normale Sup\'erieure,  Paris, France}

%\title[\texttt{achemso} Phenotypic selection]{The Effect of Phenotypic Selection on Stochastic Gene Expression}
%\begin{document}

\begin{abstract}
Genetically identical cells in the same population can take on phenotypically variable states, leading to differentiated responses to external signals, such as nutrients and drug-induced stress. Many models and experiments have focused on a description based on discrete phenotypic states. Here we consider the effects of selection acting on a single trait, which we explicitly link to the variable number of proteins expressed by a gene.
Considering different regulatory models for the gene under selection, we calculate the steady-state distribution of expression levels and show how the population adapts its expression to enhance its fitness.
We quantitatively relate the overall fitness of the population to the heritability of expression levels, and their diversity within the population. We show how selection can increase or decrease the variability in the population, alter the stability of bimodal states, and impact the switching rates between metastable attractors.

keywords: gene regulation, gene expression noise, phenotypic variability, phenotypic selection, phenotypic adaptation
\end{abstract}

\maketitle

\section{Introduction}

Within one population, individual organisms often display a large amount of observed diversity. In naturally occurring populations, some of the diversity is explained by genetic differences between the organisms. However, even in genetically identical populations, such as bacteria or yeast grown in the laboratory \cite{Fraser}, we observe phenotypic diversity, such as the variable protein levels in particular cells of the same population cultured in the same environment. This phenotypic diversity is linked to intrinsic molecular noise in gene expression stemming from relatively small copy numbers of transcription factors and the probabilistic nature of chemical reactions. While molecular noise is unavoidable, imposing physical limits to the precision of biochemical regulatory systems, it may also have a functional role \cite{Eldar}. In particular, it leads to a natural diversification of a genetically identical and otherwise homogenous population. Such cell-to-cell variability can be useful for surviving in an unexpectedly changing environment or large random fluctuations in external signals. Such arguments have been brought forward to explain the larger variable duration of competence in the native circuit of \emph{B. subtilis} than in the less noisy ``synex'' system \cite{Suel,CatagaySuel}. Another classical example is antibiotic resistance, when a fraction of bacterial cells become dormant by entering an antibiotic-resistant state without external signals, allowing the population to explore two different strategies \cite{Balabanpersistors, Balabanpersistorsmech,Balabanpersistorsrev}. In some controlled situations, phenotypic diversity was shown to underly the speed and degree of adaptation \cite{Kaneko2003,Kaneko2009}, or the capacity to switch to a more favorable phenotypic state \cite{Kashiwagi,Shimizu2011}.

Phenotypic selection under fluctuating environments has recently been studied theoretically \cite{Thattaigenetics,KusselBalabanLeibler,KusselLeibler,KusselLeibler2010, RivoireLeibler, Kaneko2006,TanasetenWolde2008}. These studies have formalized the observation that it is beneficial for populations to ``hedge their bets'' against possible environmental stresses by keeping small, specialized subpopulations able to survive in various stress conditions, at the cost of a lower fitness in normal conditions. To achieve this, cells switch stochastically between different phenotypic states, with rates adapted to the statistics of environmental changes. In this description however, the phenotypic space is usual reduced to a discrete set of states, and does not account for the molecular basis of noise.

Phenotypic differences can be directly linked to the noisy molecular nature of regulatory circuits. For example, in the competent system, small comK copy numbers are responsible for the observed noisy duration of the competent state \cite{CatagaySuel}. The large variability of gene expression is genetically encoded in the design of the circuit, for example in networks exhibiting bimodal expression \cite{Fraser}. Phenotypic variability may also take the form of ``epigenetic'' modifications, in particular on chromatin, which play an important role in eukaryotic cells. Unlike genetic variations, these different sources of phenotypic variability are not transmitted to the daughter cells in a hardwired manner. They allow populations to recover from environmental stress on much faster timescales than traditional genetic changes. As such, they allow cells to try out faster and more easily reversible strategies than genetic evolution.

The variability of protein copy numbers in monoclonal populations has been extensively studied both theoretically and experimentally \cite{Elowitz,OzbudakThattai,RaserShea,Swain,sasai_stochastic_2003}. {\comm The effect of protein concentration fluctuations on the growth rate of a genetically identical cells taking the cell cycle into account has been studied by Tanase-Nicola and ten Wolde \cite{TanasetenWolde2008}. It was shown that if the mean protein concentration is close to the value that maximizes the growth rate, fluctuations in the concentration reduce the growth rate, whereas if the mean concentration is far from the optimal, fluctuations can enhance the growth rate. A simpler continuous model of phenotypic variation under selection was studied by Sato and Kaneko \cite{Kaneko2006}.} In this paper we want to examine how selection acting on a population of genetically identical, but phenotypically variable cells, shapes the observed variability and the stability of the phenotypic states in this population. As is often done in experiments, we associate our phenotypic state with the protein copy numbers of a given type of protein. We explicitly model the stochastic dynamics of this protein, the expression of which is under the control of a simple gene regulatory network. Selection acts on the population of cells as a function of the numbers of copies of this protein. This model allows us to study the effects of phenotypic selection on observable traits in monoclonal populations.

We first study the effects of various types of selection pressures on the observed distribution in a simple model of constitutive (unregulated) gene expression. We then consider a self-activating gene, which can result in a bistable dynamical system, and we study the effect of selection on the steady-state occupancy of the two states, as well as the switching rates between them.
We also look at the effects of selection on a gene whose expression state changes on slow timescales compared to the timescale for protein change, which results in a bimodal distribution of protein copy numbers.

\begin{figure}
\includegraphics[width=\linewidth]{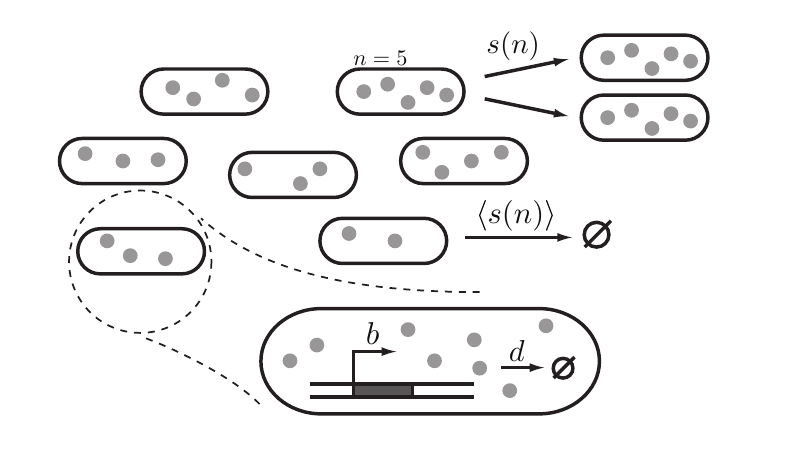}
\caption{Model of phenotypic selection by a single gene. In each cell, the number of protein $n$ undergoes a birth-death process describing the synthesis (with rate $b$) and degradation (with rate $d$) of proteins. With rate $s(n)$, cells divide. In order to keep the population size constant, the new offspring displaces another cell picked at random, creating an global and uniform death rate $\<s(n)\>$.
\label{fig:cartoon}
}
\end{figure}

\section{Model of phenotypic selection}
We assume that selection acts on a single trait---the concentration or number of copies of a given protein in the cell---denoted by $n$. 
The individual fitness of cells is defined by the $n$-dependent growth rate $s(n)$.
Within each cell, we consider the explicit dynamics of the gene expression network that produces the proteins governing the fitness of the cell. For simplicity of exposition we first assume that the gene producing these proteins is constitutively expressed. We reason directly at the level of proteins by assuming that the dynamics of mRNAs is fast. The generalization of our framework to more complicated modes of gene expression is straightforward, and we will later go beyond constitutive expression to model self-regulation.

We describe the population by the mean number of cells $\rho_n$ expressing $n$ proteins, ignoring fluctuations stemming from small numbers of cells. We will show below that this approximation works well as soon as the population is large enough.

The change in $\rho_n$ is described by a simple birth-death process accounting for the synthesis and degradation of protein molecules in each cell, and a growth rate $s_n$ experienced by each cell:
\bea
\partial_t \rho_n &=& b \rho_{n-1} +d(n+1) \rho_{n+1} -(b+dn)  \rho_{n} +s(n)  \rho_{n}\\ 
\partial_t\rho &=& {\cal L}   \rho
\label{rhoeqn1}
\eea
The growth rate is the net effect of cell division and cell death, and may be negative. Fig.~\ref{fig:cartoon} summarizes the processes governing the internal dynamics of cells as well as the population dynamics. More complex modes of regulation or gene expression dynamics can be modeled by choosing different forms for $\cal L$.

{\comm Within this model we account for the changes in the protein concentration caused by cell division by the effective degradation rate $d$, which describes the average dilution rate of proteins over a cell cycle. By doing this we do not explicitly model cell division, but we describe its consequences on the change in the protein concentration by this average rate. This is a common approach when modeling gene regulatory networks \cite{KeplerElston}, which was shown not to have a significant effect on protein concentrations (see {\em e.g.} \cite{Zwicker}). Explicitly accounting for the  effects of this punctual reduction in concentration are quite subtle and also requires accounting for the change in cellular volume. During cell division both these quantities are reduced, and since the concentration of proteins is the relevant variable for regulation, this will mostly affect the properties of the noise. As a result, in this exploratory analysis we choose to describe all dilution and degradation terms by the effective degradation rate $d$. We also neglect burst-like production effects \cite{Friedman:2006p12459,Walczak:2005p9764}, as well as the existence of an mRNA step \cite{sasai_stochastic_2003}, and replication forks affecting the birth rate \cite{Cooper}. While these effects could alter the results, analytical progress on our simplified model points the way towards more precise treatments in the future.}

Given the dynamics of the population in Eq.~\ref{rhoeqn1}, the normalized probability of finding a cell with $n$ protein copies, is given by, $p_n={\rho_n}/{\sum_{n'} \rho_{n'}}$, and follows:

\beq\label{master}
\partial_t p = ({\cal L}- \av{s}\mathbf{1}) p,
\eeq
where $\av{s}=\sum_{n} s(n) p(n)$. The addition of the selection term breaks detailed balance and introduces a nonlinearity in the master equation. A general closed-form analytical solution cannot be found for the steady state distribution. Assuming we know the value of $\av{s}$, which must be expressed in terms of the parameters of the problem, we can still write the steady state solution in the form of a series, because the problem is one dimensional. In practice we can easily find solutions numerically by iterative Euler integration. However, in certain special cases that we present below, we can find an analytical solution for the steady state distribution given $\av{s}$.

When the number of expressed proteins is large, it is useful to turn to a continuous description where the protein concentration is described by a a continuous variable $x$. Expanding Eq.~\ref{master} to second order, we get for the evolution of the probability density function $P(x)$:
\beq\label{continuous}
\begin{split}
\partial_t P(x,t) =& -\partial_x [f(x)P(x,t)] + \partial_x^2 [D(x)P(x,t)]\\
&  + (s(x)-\<s\>)P(x,t),
\end{split}
\eeq
where $f(x)=b-dx$ and $D(x)=(b+dx)/2$ are the effective ``drift'' and diffusion coefficient, respectively. $\av{s}=\int dx s(x)P(x,t)$ is the average fitness in the population.
$f(x)$ and $D(x)$ can take more general forms to account {\em e.g.} for self-regulation. This general class of models was studied in \cite{Kaneko2006} and solved in the case of a linear $s(x)$.

\section{Linear selection}\label{linear}
We first consider an exactly solvable model where the selection pressure is linearly proportional to the number of protein copies in the cells, $s(n)=s_0+s n$. 
The evolution of the mean number of proteins is given by:
\beq
\frac{d\<n\>}{dt}=b-d\<n\>+s(\<n^2\>-\<n\>^2),
\eeq
which combines the deterministic effect of birth and death with Fisher's relation \cite{Price}. 

The steady state solution of Eq.~\ref{master} can readily be found in generating function space (see Appendix \ref{appA} for details). Formally, we find an infinite family of solutions for each possible $\av{n}$, only one of which is numerically stable (stability is checked by evolving Eq.~\ref{master} iteratively using Euler's integration method). This solution is
a Poisson distribution with a rescaled mean:
\beq
\av{n}=\frac{b}{d-s}.
\eeq
At long timescales, positive selection $s>0$ acts as an effective anti-degradation term---it helps cells with large protein copy counts to survive, and eliminates cells with low copy numbers. As a result the mean of the Poisson distribution is shifted to higher protein copy numbers. Negative selection $s<0$ has the exact opposite effect.
The impact of selection on the average fitness of the population is 
\beq
\begin{split}
\<s\>&=s_0+s\<n\> = s_0+\frac{bs}{d-s}\\
&=s_1+ \frac{bs^2}{d(d-s)},
\end{split}
\eeq
where $s_1=s_0+bs/d$ is the mean value of $s(n)$ {\comm when following a single cell.}
%no {\commconcentration dependent} selection is present.
{\comm The benefit of adaptation} scales like $s^2>0$, whether selection is positive or negative, as the population adapts to find a better place in phenotypic space.

Based on this simple model, we see the general effect of selection that will come back in more complex systems. If we consider the potential landscape of the regulatory network, the system reaches a balance between the selection force $s$ that is perturbing the protein concentration in the cell, and the restoring force coefficient $d$ due to the birth-death process. For this reason, to see visible and non-trivial effects of selection, the timescales of selection and the restoring force must be comparable. For very strong selection, the mean number of protein grows uncontrollably ($\av{n}\rightarrow \infty$ when $s\to d$) as selection amplifies very rare cells with abnormally large protein numbers.
As we shall see, these effects have more visible consequences when regulation, and even more bimodality, come into play.

\begin{figure}
\includegraphics[width=\linewidth]{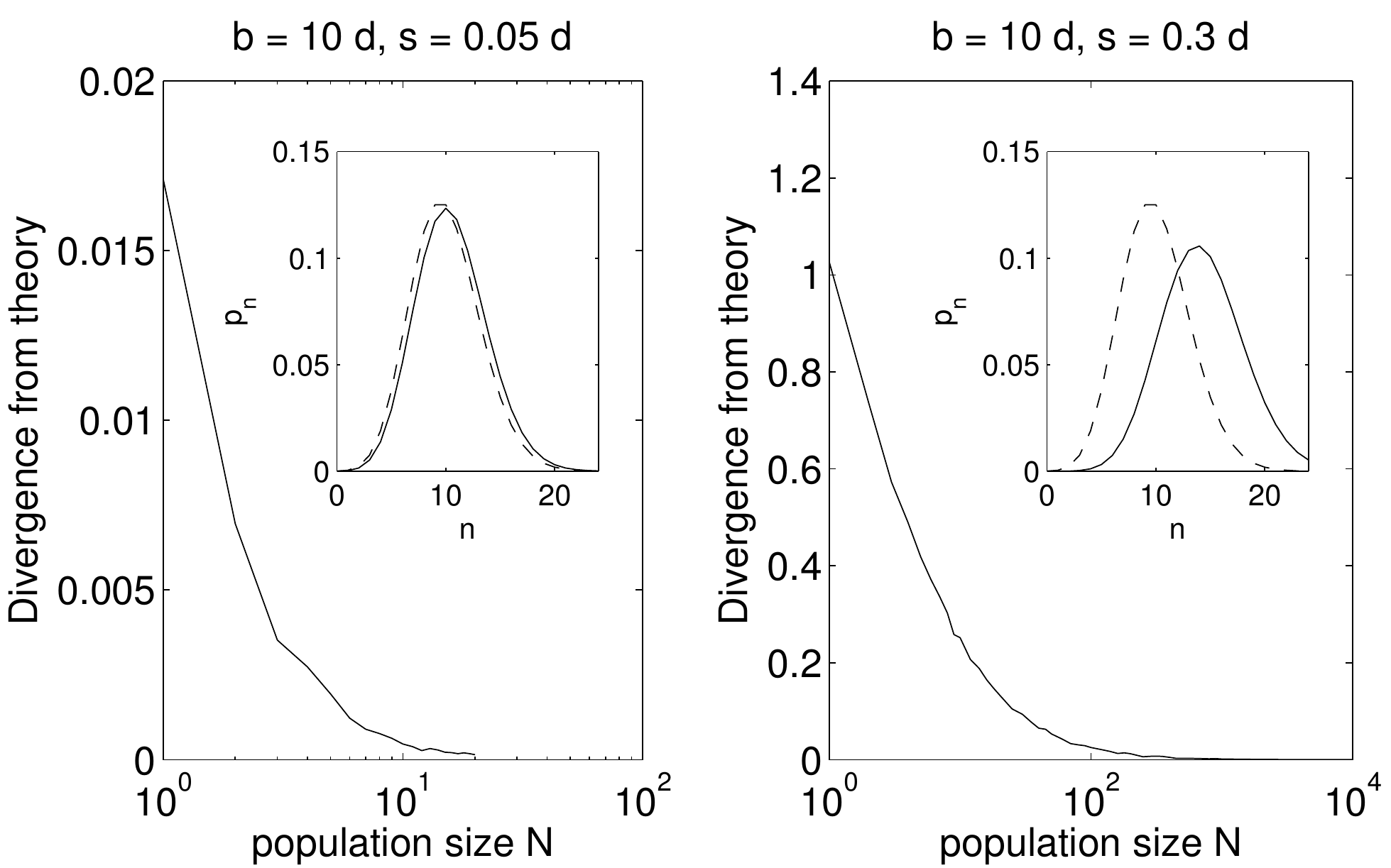}
\caption{Validity of the description by a density function. Gillespie simulations of all cells in the population are compared to the analytic prediction for $p_n$  (the fraction of cells with $n$ proteins) under linear selection ($s(n)=s_0+sn$) using the Kullback-Leibler divergence (DKL) between probability distributions. Numerical results show excellent agreement for large population sizes. {\comm Each simulation was run for a total of $10^7$ cell divisions to collect good statistics (much more than the equilibration time, which is of a few generations).}
When selection is strong ($s = 0.3 d$, right panel) larger population sizes are needed to reach good agreement than when selection is weak ($s=0.05 d$, left panel). The insets show the unselected (dashed line) and selected (full line) distributions of protein numbers.
\label{fig:DKL}
}
\end{figure}

As a general test of our mean-field approximation, whereby we reduce the system to a density function $p_n$, we verify our analytic result against Gillespie simulations of populations of cells. We explicitly consider $N$ cells, in which the gene regulatory network is modeled by a standard time varying Monte Carlo (Gillespie) algorithm \cite{gillespie, Bortz}, which appropriately models the regulation function for the different systems we consider (constitutive expression, self-activation). We assume that all cells divide stochastically with rate $s(n)$. In order to sample the steady-state distribution, we keep the population size constant, by compensating each division by the removal of a random cell. Fig.~\ref{fig:DKL} shows the difference between the analytic solution and the results of the simulation for increasing population size $N$, as measured by the Kullback-Leibler divergence (DKL). For small populations the effect of selection is moderate, and even absent in the extreme case $N=1$ where selection is irrelevent. As the population gets larger the theoretical prediction becomes more and more accurate.

\section{Self regulating gene}\label{regulation}

We now turn to study the effect of regulation on phenotypic selection. Regulation is modeled in the simplest manner by assuming that the birth rate locally depends linearly on $n$ with coefficient $b_1=\partial b/\partial n$, $b(n)\approx b+b_1n$. Let us first examine the behaviour when no selection is present. In this case the Eq.~\ref{master} can be solved using the generating function technique, and the solution reads (see Appendix \ref{appB}).
\beq\label{eq:reg}
p_n={\left(1-\frac{b_1}{d}\right)}^{b/b_1} \frac{1}{n!} {\left(\frac{b}{d}\right)}^n \prod_{i=0}^n \left(1+i\frac{b_1}{b}\right).
\eeq
Compared to the case with no regulation, the distribution is no longer Poisson: the mean shifts to $\av{n}=b/(d-b_1)$, and the Fano factor is larger: $(\<n^2\>-\<n\>^2)/\<n\>=1+\frac{b_1}{d-b_1}$.
Self-activation increases the mean and the relative variance, while self-repression decreases them both.
Solving Eq.~\ref{master} in the presence of selection, we find again an infinite family of solutions. Numerical simulations show that the only stable solution is the one that cancels one of the poles of the generating function.
This solution takes on the same functional form as Eq.~\ref{eq:reg}, where $d$ is replaced by a rescaled death rate defined as:
\beq\label{eq:regsel}
\hat d=\frac{1}{2}\left(d-s+b_1+\sqrt{(d+b_1-s)^2-4b_1d}\right),
\eeq
which simplifies to $\hat d=d-s/(1-b_1/d)$ in the limit of small selection coefficient $s$, and  to $d-s-b_1s/(d-s)$ in the limit of small $b_1$. The relative effect of selection on $\av{n}$ can be evaluated for small $s$:
\beq
\<n\>_s\approx\<n\>_{s=0}\left(1+s\frac{d}{(d-b_1)^2}\right),
\eeq
and the population fitness improvement reads $\approx s^2\<n\>_{s=0}d/(d-b_1)^2$.

The effect of positive regulation is to lower the effective restoring force to the mean value, increasing fluctuations in the protein copy number, as indicated by the increased Fano factor. These large fluctuations allow cells to explore and find regions of larger fitness, increasing the mean fitness of the population. Negative regulation has the opposite effect.

To gain further insight into this as well as other, more general models of regulation, we consider the continuous limit of the model, for which we can find an analytic solution in the small noise approximation.
Fluctuations around the steady-state value are assumed to be small. The mean steady-state concentration $x_0$ is defined by $f(x_0)=0$. In the vicinity of $x_0$, we can expand at leading order in the limit of small fluctuations: $D(x)\sim D(x_0)\equiv D$, $f(x_0)=-k(x-x_0)$ and $s(x)\approx s_0+s(x-x_0)$.
Then Eq.~\ref{continuous} simplifies to:
\beq\label{eq:FP}
\partial_t P = k\partial_x [(x-x_0)P] + D\partial_x^2 P +s(x-\av{x})P.
\eeq
The steady-state solution to $\partial_tP(x,t)=0$ is given by \cite{Kaneko2006} (see Appendix \ref{appA}):
\beq
P(x)=\frac{1}{\sqrt{2\pi D/k}}\exp\left[-\frac{k}{2D}\left(x-x_0-\frac{Ds}{k^2}\right)^2\right].
\eeq
As with the discrete birth-death process, the effect of selection is to change the mean concentration. This shift is proportional to the selection coefficient $s$, and the noise $D$. The mean population growth rate is also affected by this shift in a quadratic manner: 
\beq\label{eq:fitnessadv}
\<s\>=s_0+Ds^2/k^2.
\eeq
The parameter $k$ may physically be interpreted as the stiffness of a spring.
The larger the stiffness, the less cells are allowed to explore regions of potentially higher fitness, and the smaller the advantage confered by selection to the population. As noted in \cite{Kaneko2003,Kaneko2006}, this relation is reminiscent of the fluctuation-dissipation theorem in physics.
Within our small noise expansion, we have $b(x_0)=dx_0$, $k\approx d-b'(x_0)$ and $D\approx b(x_0)$. The parameter $b'(x_0)$ quantifies regulation and is equivalent to $b_1$ in the discrete birth death process. Activation ($b'(x_0)>0$) favors fluctuations away from the mean steady-state value, while repression $b'(x_0)<0$ suppresses them. The critical point $b'(x_0)=d$, where everything diverges, marks the transition towards a bistable system, which we discuss in Sec.~\ref{multistability}.

The scaling of the population fitness improvement (Eq.~\ref{eq:fitnessadv}) with $D$ and $k$ can be interpreted as follows. $D/k$ is the variance of protein number fluctuations, and thus quantifies the extent to which cells are allowed to explore better regions of the phenotypic space. $k^{-1}$ is the relaxation time of gene expression, and quantifies how long cells keep the memory of their internal state, and how reliably they can transmit it to their offspring across generations, {\em i.e.} their memory {\comm or} heritability. Not only is it important to hedge one's bets to adapt quickly to environmental changes, but cells must also transmit these fluctuations to offspring for the population to benefit from them in the long run. The fitness improvement due to selection is thus the product of these two features, variability {\comm($D/k$)} and heritability {\comm ($k^{-1}$)}.

\section{Threshold and cliff selection}\label{threshold}

\begin{figure*}
\includegraphics[width=\linewidth]{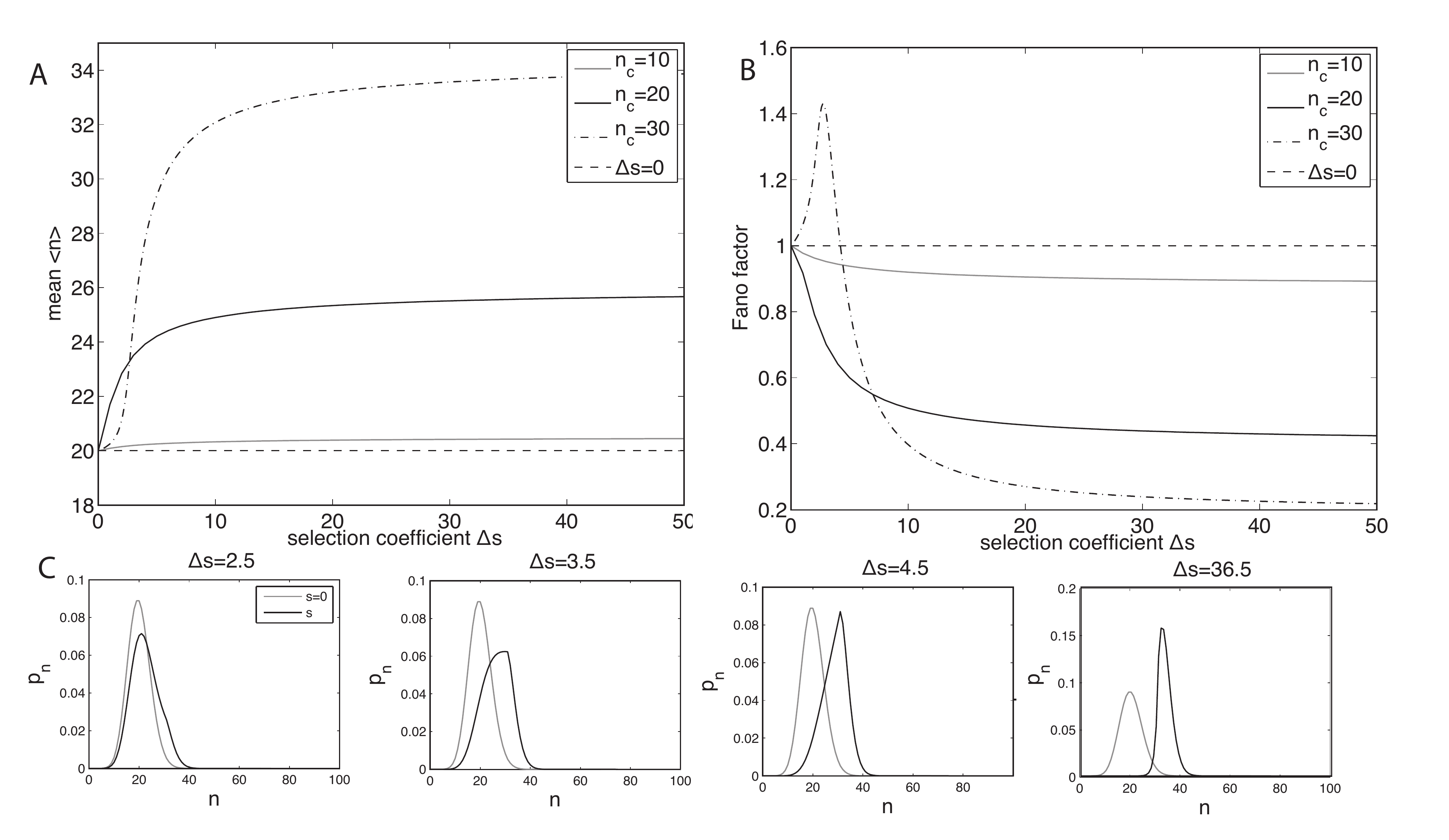}
\caption{
The effects of threshold selection pressures on the mean (A) and Fano factor, $\sigma_n^2/\av{n}$, (B) of a population of cells expressing a constitutively expressed gene. The threshold regulation function is $s=\Delta s\Theta(n-n_c)$. Different values of the position of the threshold, $n_c$, are compared to a system with no selection for a gene with $b=20$, $d=1$. Panel $C$ shows examples of the distributions compared to the Poisson distribution with $\av{n}=b/d$ that describes the system with no selection for different selection pressures for $n_c=30$. The variance of the distribution increases for small selection pressures and then the mean shifts to higher values, resulting in the initial increase and then decrease of the Fano factor.
\label{fig:thres}
}
\end{figure*}

Bacterial cells grown in the presence of an antibiotic can develop resistance to the drug without changing the genome \cite{Balabanpersistors, Balazsi, Fraser, Blake}, by expressing an antibiotic resistance gene above a given threshold.
To describe this situation we assume that cells with at least $n_c$ of protein copies reproduce  with rate $s_0$ in the presence of the drug, whereas cells with $n<n_c$ grow with rate $s_1$. The selective pressure now takes the form of a step function, $s(n)=s_1+(s_0-s_1)\Theta(n-n_c)$. We call this scenario threshold selection. Note that because of normalisation, the distribution of protein levels in the population does not depend on the absolute scale of $s(n)$, and the only relevant parameter is $\Delta s=s_0-s_1$.
In the extreme case where cells under the threshold die, we have $\Delta s=+\infty$. We call this scenario cliff selection.

By inspecting the effects of threshold selection on a constitutively expressed gene with a mean expression of $\av{n}=20$ protein copies  in the absence of selection (Fig.~\ref{fig:thres}A), we see that even moderate selection pressures acting within the variance of the mean of the distribution result in a steep increase in the mean number of proteins. The cells that express small numbers of proteins now have a fitness disadvantage and hence the distribution shifts to have a higher fraction of cells expressing more proteins. For even larger selection pressures, the cells with $n\geq n_c$ are favoured, but the mean production rate in the cells remains the same. Therefore a balance is reached between the restoring force due to protein degradation, which brings protein copy counts in cells down below the threshold to $n<n_c$ hindering their reproduction, and the proliferation of cells with $n\geq n_c$. 
The mean number of proteins in a population thus reaches a plateau for the cliff model, $\Delta s\to +\infty$.

When $\Delta s$ is large, as the mean number of proteins increases with selection pressure, the variance decreases and the distribution becomes subpoissonian as shown by the decrease in the Fano factor (see Fig.~\ref{fig:thres}B). 
The variability of the population is thus reduced.
Cells that survive the selection pressure have more offspring and effectively transmit information about their expression state to the next generation, while cells that produce less than the threshold are less likely to have offspring. However for a relatively large critical value of $n_c$ (dash-dotted line in  Fig.~\ref{fig:thres}B), the Fano factor increases for small selection pressures. In this case a small fraction of cells are to the right of the threshold and bear a selective advantage. When this advantage is still small, this causes the tail of the distribution to get slightly fatter, thus widening the distribution, but without significantly affecting the mean. For larger selection pressures, the advantage of expressing more proteins becomes significant and we observe a cusp in the probability distribution at $n=n_c$ (Fig.~\ref{fig:thres}C).

\begin{figure}
\includegraphics[width=\linewidth]{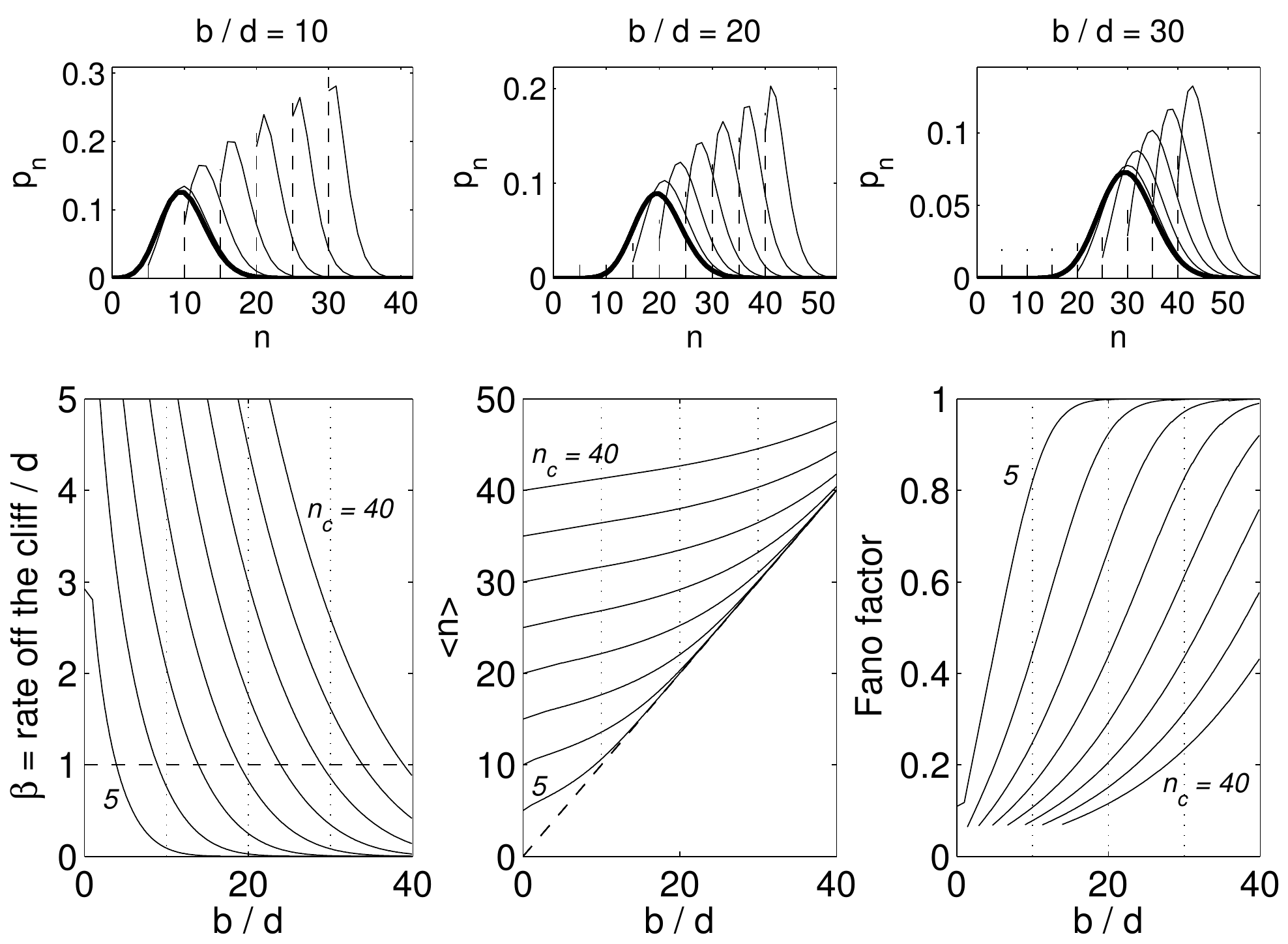}
\caption{Effect of cliff selection on the population. Bottom, from left to right: cell death rate, mean protein level, and Fano factor (variance over the mean) as a function of the mean unselected expression level $b/d$, for various values of $n_c$. Selection increases the mean protein level above $n_c$. For high thresholds, cells cluster around $n_c$, resulting in low variances as shown by the Fano factor.
This effect becomes smaller when the unselected expression is large compared to the threshold, $b/d\gg n_c$. In the leftmost plot, for a given proliferation rate $s_0$, $\beta=s_0/d$ marks the transition between extinction and proliferation. The lines can thus be interpreted as separatrices between these two phases in the space $(b/d,s_0/d)$ for various values of $n_c$. For $\beta>1$ (dashed line), regulation is needed to achieve values of $s_0/d$ that ensure survival.
 Top: example distributions of protein numbers $p_n$ for the values of unselected expression levels $b/d$ marked by dotted line in the bottom plots. The dashed line shows the location of the threshold.
\label{fig:cliff}
}
\end{figure}

In the limit case of cliff selection $\Delta s\to +\infty$, where the effect of drugs is most detrimental, one can solve formally for steady state via the generating function (see Appendix \ref{appC}). As before we find a family of solutions $p^{(\beta')}_n$, parametrized by a single number $\beta' =n_cp_{n_c}$. $\beta'$ must be smaller than some critical $\beta$, defined such that for $\beta'>\beta$, $p^{(\beta)}_n$ becomes negative, making the solution unphysical.
Numerical stability analysis shows that the only stable solution is in fact found at this critical value $\beta$. 
The average growth rate in the population can be calculated from Eq.~\ref{master} and its value is $s_0-\beta d$: the proliferation of cells $s_0$, minus the flux of cells falling off the cliff, $\beta d$.
Therefore $\beta$ is the rate of cell death in units of the degradation rate. It is shown as a function of $b/d$ for several values of $n_c$ in Fig.~\ref{fig:cliff}.

The value $\beta=s_0/d$ marks the transition between the two phases of population: extinction and proliferation. When $\beta>s_0/d$ (extinction), the lifespan of the population under stress is given by $\log(N_0)/(\beta d-s_0)$, where $N_0$ is the initial population size.
Note that biologically, in the absence of regulation, the death rate should be larger than the division rate because of dilution, $s_0/d\leq 1$. $\beta=1$ therefore represents a best case scenario where degradation is kept to a minimum, and survival is maximum. The average protein level is given by $\av{n}=[b/d-\beta(n_c-1)]/[1-\beta]$. Therefore the transition at $\beta=1$ is obtained at $b/d=n_c-1$.

We have seen that the effect of regulation was to rescale $d$ to $d-\partial b/\partial n$, making it possible to have $s_0/d>1$. In that case, the transition between extinction and proliferation is reached at higher $\beta$, and therefore at smaller mean expression levels $b/d$ (see Fig.~\ref{fig:cliff}, bottom left). In other words, positive regulation and the concomittent increased variability allow the population to better survive an acute stress.

The continuous couterpart of the cliff model can also be solved, with $f(x)=-kx$, $D(x)=D$, and $s=s_0$ for $x>x_c$, and $-\infty$ otherwise. As in the discrete case, there exists a $\beta_c$ above which $P(x)$ becomes unphysical. Numerical experiments show that this $\beta_c$ is the only stable solution. The average growth rate of the population is $s_0-\beta D$. $\beta=s_0/D$ gives the boundary in phase space between extinction and proliferation. The average concentration is given by $\av{x}=(1-k/\beta D)^{-1}x_c$, indicating that $\beta=k/D$ when $x_c=0$. At that particular point the solution is simply $P(x)=kx/D \exp(-kx^2/2D)$.

\section{Multistability}\label{multistability}

In Sec.~\ref{regulation} we have discussed the importance of the heritability of the expressed number of proteins for the population to benefit from selection.
One of the mechanisms that has been proposed \cite{Weinberger, Fraser} to stabilize phenotypic states of cells with higher fitness is self-activation of genes. In a large parameter regime self-activating gene circuits are bistable. There are two deterministic steady state expression states: one with a high number of protein copies and one with a low number of protein copies. Self-activation stabilizes these two states and leads to two stable subpopulations, allowing the population of cells to respond to different pressures. This simple scenario has been studied extensively in the literature \cite{Thattaigenetics,KusselBalabanLeibler,Acar,Balazsi}. Here we consider the effects of selection on the diversity of the responses of the population, and the stability of each of these states, within a concrete model of gene expression that displays bistability through a steep self-regulating function, $b(n)=(b_0K^2+b_1n^2)(K^2+n^2)$.

\begin{figure*}
\includegraphics[width=.7\linewidth]{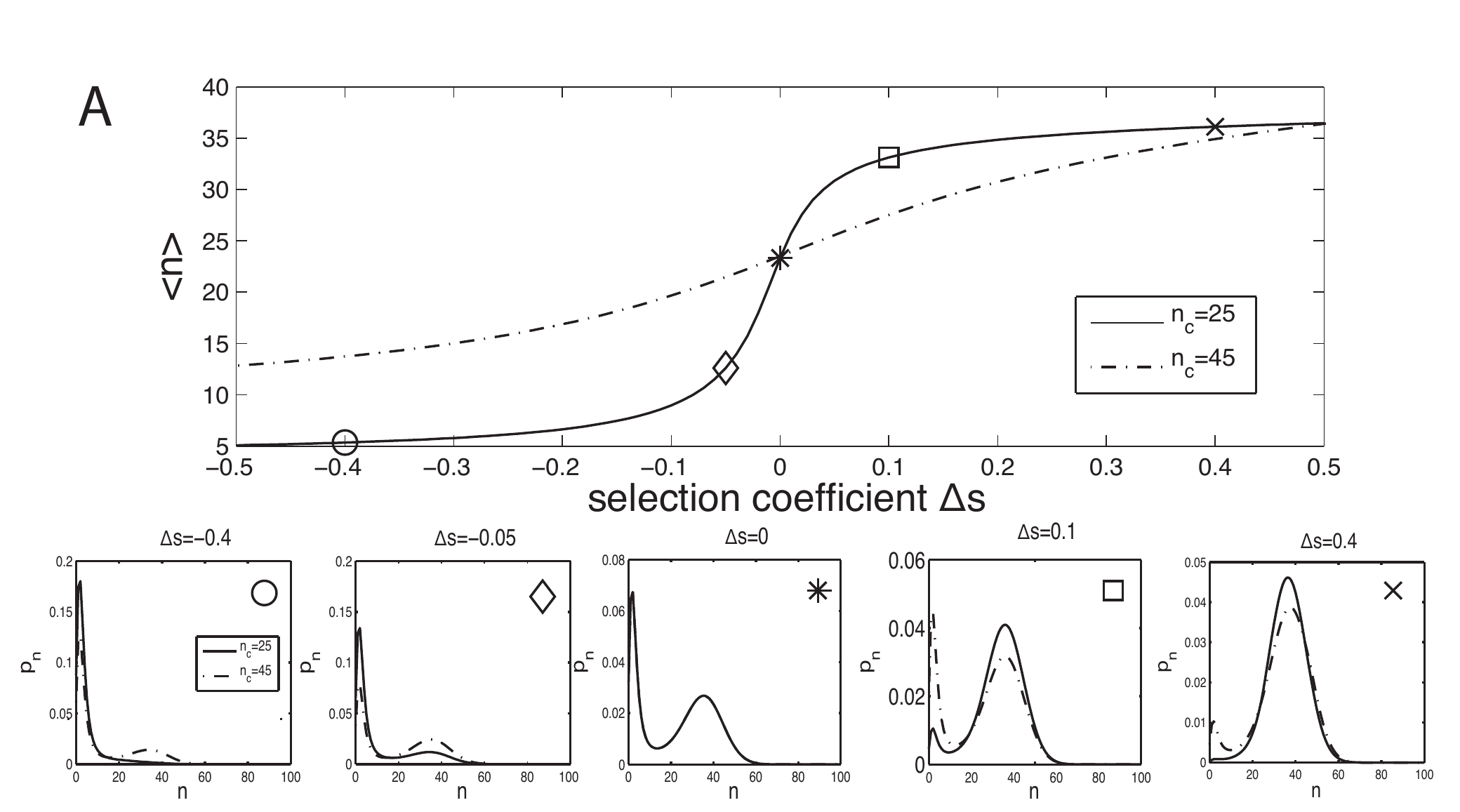}
\includegraphics[width=.7\linewidth]{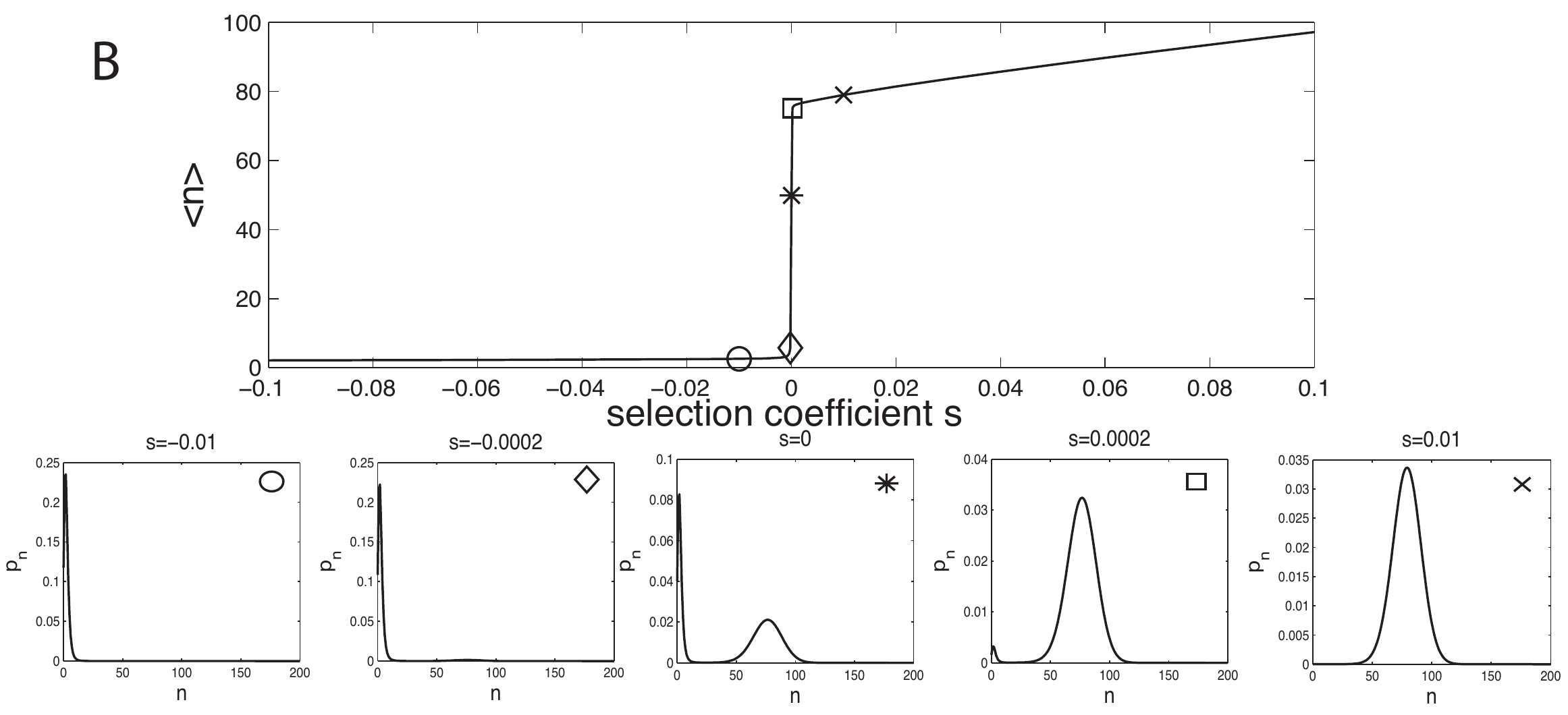}
\caption{The effects of selection on a population of self-activating bistable genes in the case of threshold ($s(n)=s_1+\Delta s\Theta(n-n_c)$, panel A) and linear ($s(n)=s_0+sn$, panel B) selection. The mean number of proteins is shown as a function of the selection coefficient. Probability distributions for indicated values of the selection coefficient are plotted in the bottom of each panel. Panel $A$ shows a comparison between two critical values of proteins, $n_c=25$ which falls between the two expression states (solid line) and $n_c=45>\av{n}$. Regulation parameters were chosen to have nearly equal probability to be in the high and low expression states in the absence of selection, $\Delta s,\,s=0$. In both cases $b(n)=\frac{b_0K^2+b_1n^2}{K^2+n^2}$. For threshold regulation: $b_0=2$, $b_1=50$, $d=1$, $K=22.5$. For linear regulation: $b_0=2$, $b_1=100$, $d=1$, $K=42$. This value of $K$ for the linear case was chosen to ensure slow switching between the two states. For smaller $K$ the change in $\av{n}$ as a function of the selection coefficient is even sharper.
\label{fig:switch}
}
\end{figure*}

Within the models of selection we have discussed so far, the high protein number state is favoured by selection. In Fig.~\ref{fig:switch} we show the effects of threshold (Fig.~\ref{fig:switch} $A$) and linear (Fig.~\ref{fig:switch} $B$) selection on the mean number of proteins in a population of cells with bistable genes. These genes have a close-to-equal probability of expressing proteins in high and low numbers in the absence of selection. As discussed earlier, cells that express low protein copy numbers are less likely to reproduce when selection is present. However bistability greatly amplifies this difference.
For positive selection, the low protein copy expression state is virtually eliminated and the {\comm population looses its bimodal nature}, and hence its diversity. Analogously, negative selection pressures eliminate the high protein copy number expression state. This is illustrated by the probability distributions of protein expression shown in Fig.~\ref{fig:switch}.

Unlike in the case of the constitutively expressed (unregulated) gene, where the effects of linear selection pressure were quite smooth ($\av{n}={b}/{(d-s)}$), the bistable expression results in a population that is very susceptible to selection and a steep transition in the mean number of expressed proteins. Selection effectively acts on the expression states (low/high) and does not discriminate cells that differ by a few numbers of proteins, resulting in a threshold response for linear selection. For this reason, the behaviour is not much affected by the precise form of the selection function.
For large, linear selection pressures, the distribution becomes unimodal and then we recover the same behaviour as in the unregulated gene discussed in section \ref{linear}---an increase in the mean number of proteins as $\av{n}={b}/{(d-s)}$. 
In the case of threshold selection, for large selection pressures (positive or negative) the system also behaves effectively like an unregulated gene, and the mean number of proteins reaches a plateau.

Large threshold values $n_c$ stabilize the expression in the low state, as illustrated by the dashed lines of Fig.~\ref{fig:switch}A. Similarly to the case of no regulation discussed in Sec.~\ref{threshold}, the fraction of cells above $n_c$ is low for moderate $\Delta s$, resulting in a less fit but more diverse population than for lower values of $n_c$. 

As the mean number of proteins expressed by the genes increases, the response of the system to selection becomes steeper. The mean number of proteins expressed in the population in Fig.~\ref{fig:switch}B is roughly double that expressed in the population in Fig.~\ref{fig:switch}A. The noise in the latter system is higher, resulting in more frequent switching between the low and high expression states.
which can be seen by comparing the height of the barrier between the two states in the probability distributions in the absence of selection.
As a result, low noise further amplifies the effects of selection by freezing the expression state in the lifetime of a cell, thus increasing the heritability of its state. 

In the limit of small noise, the system can thus be reduced to two states: low or high expression. If we know the transition rates $k_+$ and $k_-$ from low to high and from high to low, as well as the average selection coefficient $s_-=\sum_{n<n_0} s(n)p_n$ and $s_+=\sum_{n>n_0} s(n)p_n$ in the low and high states  (where $n_0$ is the midpoint between the two states), we can write coupled equations for the number of cells in each of the two states, $\rho_+=\sum_{n>n_0} p_n$ and $\rho_-=\sum_{n\leq n_0}p_n$:
\bea\label{bistable}
\frac{d\rho_+}{dt}&=&k_+\rho_- - k_-\rho_++s_+\rho_+,\\
\frac{d\rho_-}{dt}&=&k_-\rho_+-k_+\rho_-+s_-\rho_-.
\eea
These equations are commonly used to describe growing populations with two states \cite{Thattaigenetics}. They have been proposed in the context of bacterial persistence \cite{KusselBalabanLeibler} to model the switching between normal and persister cells in {\em E. coli}, or betwen the low and high expression states of a antibiotic resistance gene in {\em S. cerevisiae} \cite{Balazsi}, and has also been used in the context of the galactose utilization network of {\em S. cerevisiae} \cite{Acar}.
These equations can be readily solved at steady state, yielding the fraction of cells in the high state,
\beq\label{eq:twostate}
p_+=\frac{\rho_+}{\rho_++\rho_-}=\frac{\sqrt{(\bar{k}-\Delta s)^2+k_+\Delta s}-\bar k+\Delta s}{2\Delta s},
\eeq
where $\bar k=k_++k_-$ and $\Delta s=s_+-s_-$.
When selection is negligible compared to the switching rate, $\Delta s\ll k$, one recovers the equilibrium occupancy of a two-state model: $p_+=k_+/\bar k$. In the opposite limit, $k\ll \Delta s$, where switching is rare, cells that are in the most favorable of the two states will proliferate and outcompete cells from the other state, and will do so much faster than they switch between the two states: $p_+=(1/2)(1+\mathrm{sign}(\Delta s)$. This describes well the situation shown in Fig.~\ref{fig:switch}:
cells in the bistable population lose their diversity and all express high (low) numbers of proteins when selection is positive (negative).

\section{Switching rate between metastable states}

We have seen that selection could destabilise metastable states, especially when switching is very rare compared to the differences in growth rate. In that case, if we assume that the whole population is prepared in the state of lowest fitness, it typically takes only one cell to make the transition, in order for the whole population to follow suit and switch. Once that first cell has switched, it proliferates and its offspring quickly outcompete the cells that have remained in the state of lower fitness. {\comm This implies that large populations are more likely to adapt rapidly because of their increased chance of switching, as confirmed experimentally by Shimizu {\em et al.} \cite{Shimizu2011}.}
When switching is even so rare that it is unlikely for a single cell out of a very large population to switch, $Nk_+\ll 1$, selection could have another, more subtle effect on the switching rate itself, by enhancing (or suppressing) the rare trajectories in gene expression space that make the transition. Cells that explore rare events towards the separatrix between the two states may be rewarded (or punished) by being allowed to reproduce (or made to die), therefore increasing (or decreasing) the future chance for a cell or its offspring to make the transition.

\begin{figure}[h!]
\begin{center}
\includegraphics[width=\linewidth]{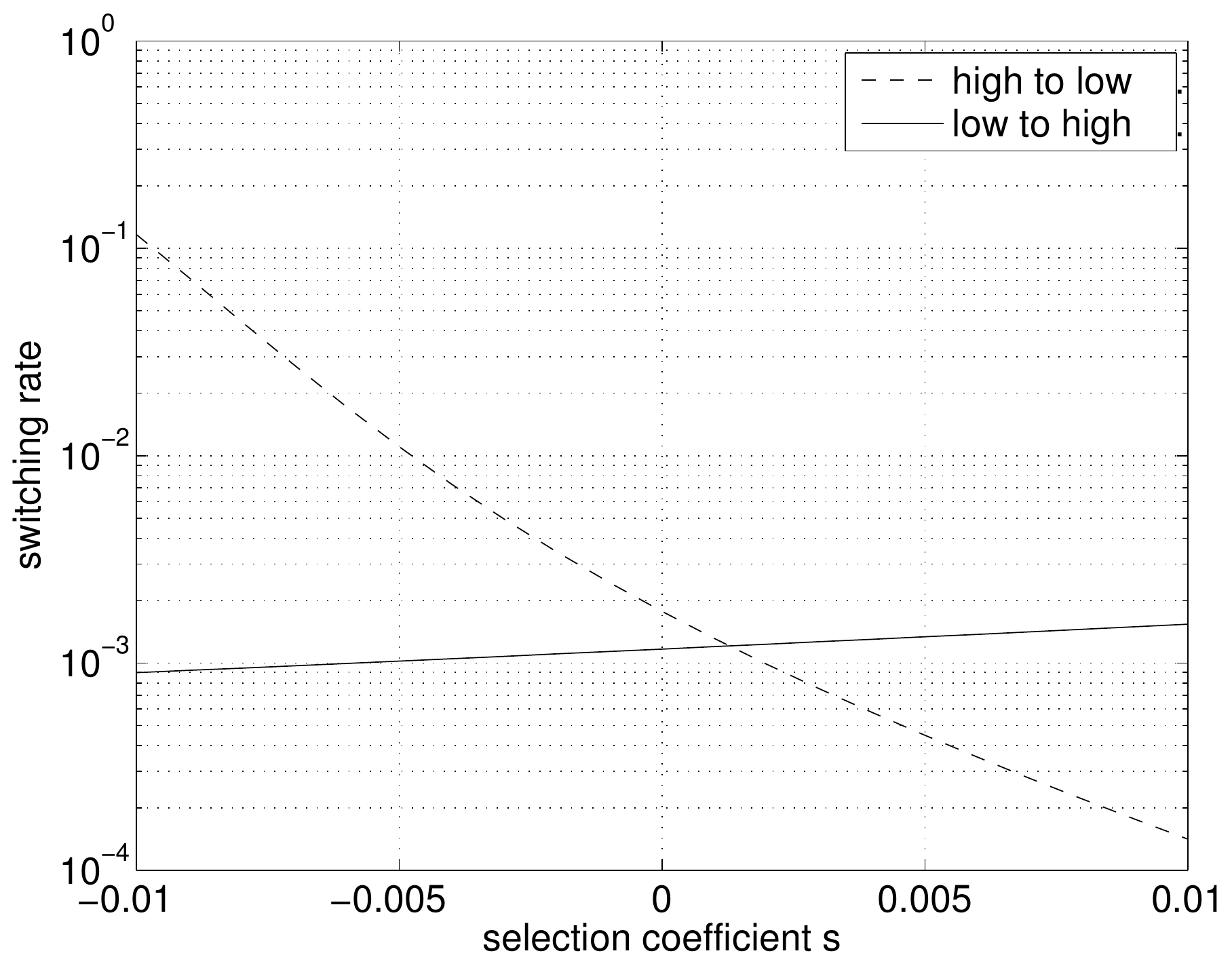}
\caption{Rates between the low and high states in the model of a self-regulating gene discussed in Fig.~\ref{fig:switch}B, as a function of the linear selection coefficient. 
\label{fig:rates}
}
\end{center}
\end{figure}

In practical terms, we would like to calculate the probability that a single cell, or one of its offspring, escape the basin of attraction of a given state. This is a slightly different problem than the one we are faced with when dealing with a homogenous population of cells that is not under selection, because selection breaks detailed balance and favours some cells over others. Because of this, traditional mean first passage methods are not applicable. However we can calculate these rates by solving Eq.~\ref{master} conditioned on cells not switching, which is implemented by a reflecting boundary condition at the midpoint $n_0$ between the two states. By computing the rate of cells that would go through $n_0$, we obtain a numerical estimate for the rate of first passage of a single cell. Fig.~\ref{fig:rates} shows the rates between the low and high states in the self-regulating bistable gene discussed in Fig.~\ref{fig:switch}B, as a function of the linear selection coefficient $s$. The effect of selection is to enhance transitions from the unfavorable to the favorable state by giving a selective advantage to cells that venture towards the transition point.

\begin{figure}[h!]
\begin{center}
\includegraphics[width=.5\linewidth]{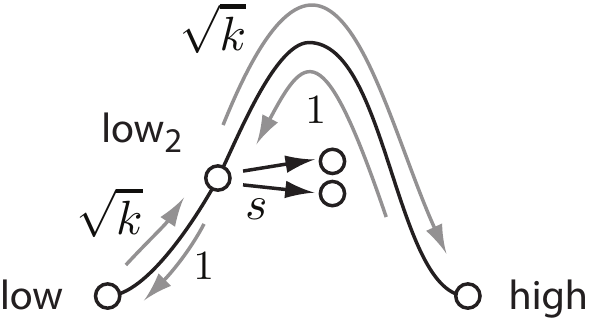}
\caption{A toy model for selection-aided switching. Cells transition from the low to high states via an intermediate state low${}_2$, in which they are allowed to reproduce with rate $s$.
\label{fig:toy}
}
\end{center}
\end{figure}

To better understand this enhancement, we first consider a simplified version of the problem, where there are only three effective states: low, high, and an intermediate state low${}_2$ between the low state and the transition point between the high and low states (see Fig.~\ref{fig:toy}). The transition rate from low to low${}_2$ is $\sqrt{k}$, and from low${}_2$ to high is $\sqrt{k}$. Time is rescaled so that the transition rate from low${}_2$ to low is set to 1. 
The selective advantage (or disadvantage) along the reaction path is modeled by setting the growth rate to $0$ in the low state, and $s$ in the low${}_2$ state.
The population maintains a constant population size $N$. The transition rates are very low, so that $\sqrt{k}\ll1 $ and $kN\ll 1$.
Starting with all cells in the low state, we ask how long it take for at least one cell to transition to high. Before the transition happens, the system is described by the number of cells in the low and low${}_2$ states, $N_1$ and $N_2$, with $N=N_1+N_2$. We treat all states where at least one cell made it to the high state as one big absorbing state. After some relaxation time, the rate of escape into the absorbing state is given by $\sqrt{k}\<n_2\>$, where $\<n_2\>$ is the average number of cells in the low${}_{2}$ state at quasi-equilibrium ({\em cf.} Eq.~\ref{eq:twostate} with $k_+=\sqrt{k}$ and $k_-=1$), and is given by $\sqrt{k}N/(1-s)$. The rate of passage of the first cell to the high state is given by:
\beq
\frac{kN}{1-s}.
\eeq
As $s\to 1$, cells in state low${}_2$ reproduce almost as fast as they switch back to low, providing an increasing chance for switching to the high state.

This first passage problem can also be studied within the small noise approximation. In this limit, the number of protein copies $x$ follows a random walk with drift $f(x)$ and diffusion coefficient $D(x)$ under a selection coefficient $s(x)$ (see Appendix \ref{appD}). In the limit $D(x)\to 0$, the optimal reaction path can be calculated and satisfies: $dx/dt=\pm \sqrt{f(x)^2-4D(x)(s(x)-\<s\>)}$, where $\<s\>$ is the average fitness of the population in the basin of attraction. The switching rate is given by the action of the optimal path, $\sim \exp(\mathcal{A})$.
In the limit of small noise, this action reads: 
\beq\label{enh}
\mathcal{A}=\mathcal{A}_0+\int_{x_{\rm initial}}^{x_{\rm final}} dx\, {[s(x)-\av{s}]}/{|f(x)|},
\eeq
where $\mathcal{A}_0$ is the action in absence of selection. When going against a constant drift $f(x)$, the enhancement of the rate is just proportional to the mean selective advantage along the path. The stronger the adverse drift $f(x)$, the smaller the enhancement. The rarity of switching is typically affected by two factors: the strength of the adverse drift, and the distance to the transition point in phenotypic space. The enhancement of Eq.~\ref{enh} is expected to have a strong effect on transitions limited by long distances to the transition point and weak adverse drifts, and only a moderate effect on transitions limited by strong adverse drifts over short distances. This explains the difference between the impact of selection on the two rates between the high and low states in Fig.~\ref{fig:rates}. Although the two rates are comparable in the absence of selection, the transition point $n_0\approx 19$ is much closer to the low state ($n\approx 2$) than to the high state ($n\approx 100$), and therefore is less impacted by selection.

Another interesting case is that of a constant stiffness, $f(x)=-k(x-x_0)$, and linear selection $s(x)=s_0+sx$. For small $s$ we have $\<s\>\approx s(x_0)$, and we get: 
\beq
\mathcal{A}=\mathcal{A}_0 + \frac{s(x_{\rm final})- s(x_{\rm initial})}{k}.
\eeq
In this case the improvement in the switching rate is simply proportional to the fitness difference between the initial and final states.

Taken together, these different estimates indicate that selective pressure has a significant (${\cal O}(\Delta s)$) effect on the rate of passage of the first cell. This is however a rather moderate effect compared to that on the steady-state occupancy of the metastable states (Eq.~\ref{eq:twostate}).

\section{Non-adiabatic model}
So far we have assumed that the binding and unbinding of any regulatory molecules occurs on very fast timescales compared to the timescale on which the protein number changes. Experiments have shown that in the case of many systems the change of the gene expression state \cite{golding_eukaryotic_2006,idopaulssonbact, caixie, chubb_transcriptional_2006} (from enhanced to basal expression and vice versa) can occur on timescales comparable with those on which the protein number changes. These types of models have been shown to result in a bimodal steady state distribution of protein numbers \cite{walczak_self-consistent_2005, hornos, walczakmuglerwigginsreview}, where one peak corresponds to protein expression when the gene is in the enhanced state and the other when the gene is the basal state. In this case, since the protein number and gene states change on comparable timescales, the protein number state can equilibrate in each of the gene states before it changes. Although the detailed positions of these two peaks depend on the type of regulatory model (self-activation, self-repression, regulation by an external transcription factor protein), the general properties do not depend on the details of the binding rate. Therefore for simplicity of exposition we choose to present the problem for a gene that is regulated by an external transcription factor, resulting in a constant binding rate $\omega_+$. We then discuss the results for self-activation when the transcription factor protein binds as a dimer, $\omega_+=hn^2/2$, where $h$ is the binding rate coefficient.

Specifically, we consider the joint probability that the gene is in the enhanced ($+$) or basal ($-$) expression state, and that $n$ copies of the protein are present in the cell. Formally we have two density functions $(\rho^-_n,\rho^+_n)$, and their associated normalized densities $(p^-_n,p^+_n)$ with $\sum_n (p^-_n + p^+_n)=1$. We can then write down the dynamics of this system as an extended birth death process, which also accounts for binding and unbinding of the activating protein:
\bea
\partial_t \rho^-_n &=& \sum_{n'}{\cal L}^{BD,-}_{n,n'}\rho^-_{n'}+ [s(n) - \omega_+] \rho^-_n +\omega_- \rho^+_n \\
\partial_t \rho^+_n &=& \sum_{n'}{\cal L}^{BD,+}_{n,n'}\rho^+_{n'}  + [s(n) - \omega_-] \rho^+_n +\omega_+ \rho^-_n,
\label{nonad}
\eea
where ${\cal L}^{BD,\pm}$ are the birth-death operators describing protein synthesis in the enhanced or basal gene expression state, and $\omega_+$ and $\omega_-$ are the binding and unbinding rates of the transcription factor.

\begin{figure}
\includegraphics[width=\linewidth]{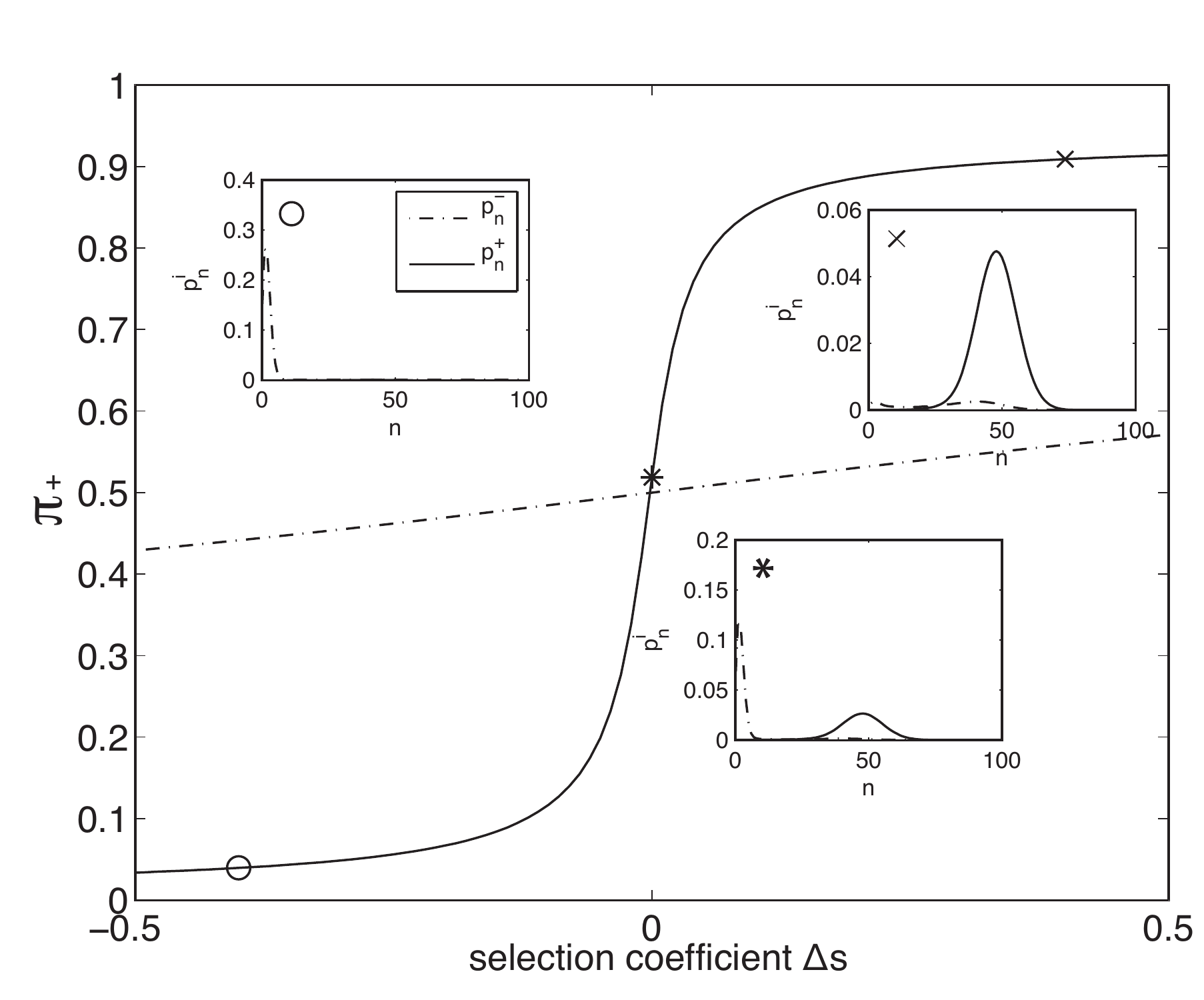}
\caption{The effects of selection on a population of nonadiabatic genes,slowly transitioning between on and off states. The probability for the gene to be found in the enhanced expression state, $\pi_+$, is shown as a function of the selection pressure for threshold selection pressures acting on constitutive genes, $\omega_+=\omega_-/K$, (dashed line) and self-activating genes with dimers binding, $\omega_+=hn^2/2$ (solid line). The inserts show examples of the probability distributions for $s=-0.4$ (circle), $s=0$ (asterix) and $s=0.4$ (cross) for the self-activating gene. The threshold is taken at $n_c=25$, with $b_-=2$, $b_+=50$, $d=1$, $\omega_-=0.5$. $K=1$ for the constitutive gene and $h=\omega_-/K$, $K=7.7$ for the self-activating gene. We note that in the adiabatic regime ($\omega\gg 1$), $\pi_+=0.5$ for all $s_0$ for the constitutive gene, however $\av{n}$ changes (see Fig. 3). The self-activating gene in the adiabatic regime is discussed in Fig. 6.
\label{fig:nonad}
}
\end{figure}

When selection is linear, $s(n)=s_0+sn$, an analytical solution to the steady state distribution can be found in generating function space  \cite{szymanska, hornos, walczakmuglerwigginsreview} in terms of Whittaker functions \cite{table}, given that we know the mean number of protein copies in the system, similarly to the previously discussed systems.

More intuition about the effects of selection can be gained from the fraction of cells that have genes in the enhanced state, $\pi_+=\sum_n p^+_n$, which is shown as a function of selection in Fig.~\ref{nonad} for the unregulated gene and the self-activated gene, assuming transcription factors bind as dimers ($\omega_+=hn^2/2$) and threshold selection $s(n)=s_0+\Delta s\Theta(n-n_c)$.  An analysis of an effective two state system similar to the one presented in section~\ref{multistability} (Eq.~\ref{bistable}) can help us understand the probability for the gene to be expressed at an enhanced rate, $\pi_+$, for the constitutive gene. Summing Eq.~\ref{nonad} over the number of protein copies and solving for $\pi_+$, we obtain:
\beq
\pi_+=\frac{\omega_++\Delta s\sum_{n>n_c} p^+_n}{\omega_++\omega_-+\Delta s\sum_{n>n_c} (p^+_n+p^-_n)}.
\eeq

As $\Delta s\rightarrow 0$ we recover the equilibrium result of the binding and unbinding rates, $\pi_+=\frac{\omega_+}{\omega_++\omega_-}$. For large selection pressures compared to the binding/unbinding rates, $\pi_+=\frac{\sum_{n>n_c} p^+_n}{\sum_{n>n_c} (p^+_n+p^-_n)}$ is given by the fraction of cells that have more proteins than the threshold and their genes are in the enhanced expression state and tends to $1$ for large $\Delta s$. Similarly, for negative selection pressures, ($\Delta s<0$), $\pi_-$ tends to 1 for large negative $s_0$.

This behaviour is shown in Fig.~\ref{fig:nonad}. We choose parameters for which the probability of the gene to be expressed in the enhanced and basal state is equal in the absence of selection.
Selecting for a large number of proteins favours cells that are in the enhanced state and vice-versa. This effect, already visible for constitutive expression, is made more pronounced when feedback is present.  
Examples of distributions of the fraction of cells that  have $n$ protein copies and the gene is in the enhanced ($p^+_n$) or basal state ($p^-_n$) are plotted for different selection coefficients in the case of threshold regulation for a self-activating gene, assuming transcription factors bind as dimers ($\omega_+=hn^2/2$). The change in the distributions are qualitatively similar for the constitutive gene. We explicitly see that strong positive selection favours the enhanced state.

In summary, as in the case of abiabatic regulation discussed in Sec.~\ref{multistability},
selection destroys the one of the modes in bimodal systems, reducing the observed variability, even in the absence of regulation. 
This effect is expected to be stronger as the binding/unbinding rate is smaller, and is strongly amplified by positive regulation.

\section{Conclusion}
We have shown how selection acting on a simple phenotypic trait such as the expression level of a gene, could significantly affect its mean expression level, diversity, and stability, to the benefit of the population of cells as a whole.

The adaptation of monoclonal populations to challenging environmental conditions, such as antibiotic stress or nutrition shortages, as studied experimentally in yeast \cite{Balazsi, Acar} and {\em E. coli} \cite{KusselBalabanLeibler,Kashiwagi,Shimizu2011}, is usually described by models of switching between a finite number of states.
Our approach goes beyond this coarse-grained description, and studies the effects of selection on the full spectrum of expression levels. In particular, we have characterized the
stability and variability of expression within a single metastable state, within a simple model of constitutive expression. In this case, in the small noise approximation, Eq.~\ref{eq:fitnessadv} quantifies how the population improves its overall fitness proportionally to the heritability $k^{-1}$ and the variability $D/k$ of fluctuations in protein copies. Heritability can be enhanced by means of positive regulation, which decreases the relaxation rate $k$.

When regulation is strong $k<0$, the system can become bistable, with two states of low and high expression level. This is a case of very strong heritability, in which cells can transmit their expression state to their offspring over many generations.
We have shown that selection destroys the bimodality of the distribution of gene expression, by favoring the state of highest fitness. This effect is all the more important when differences in growth rate between the states are large compared to the switching rates. {\comm The phenomenon provides a simple response system at the population level, driven by the proliferation of the fittest cells rather than by direct cues from a signaling pathway. The relative importance of this adaptive response, compared to signaling, was assessed experimentally and discussed in \cite{Kashiwagi} for a synthetic toggle switch system in {\em E. coli}. In particular, it was shown that the adaptive response was sufficient to observe reliable switching to the state of higher fitness.}

In multistable systems, selection decreases the variability of a population by favoring some metastable states over others. However, within a single metastable state, a linear selection in the expression level mostly affects the mean and stability of expression, but not its variance. By contrast, when selection is step-like, with a different growth rate below or above a given threshold of expression, selection may increase or decrease variability, depending on the strength of selection. Very stringent selection tends to decrease the diversity of  expression at the cost of fitness (Fig.~\ref{fig:cliff}), while a moderate selection acting on the tail the distribution increase the variance by amplifying these tails (Fig.~\ref{fig:thres}). 

In bistable systems, selection has another overlooked effect, which cannot be grasped by a simple two-state model: it enhances or suppresses the rate of switching between the two states, by giving a selective (dis)advantage to cells going along the transition path. This selection-aided switching could serve as a mechanism for driving and stabilizing a population of cells through differentiation using a gradual selective pressure, for example during developement where phenotypic noise plays an important role \cite{losickdesplan}.

Our results show that selective pressure acting on the expression of a single gene may strongly affect its behaviour at the population level. It would be interesting to test this idea experimentally, by measuring the properties of gene expression (mean, variance, switching rates) in selective against non selective environments, for different modes and strength of regulation. For example such experiments could test the prediction that positive regulation enhances the effect of selection on the population mean.

Our approach provides a broad framework for addressing the effect of selection on observable phenotypic traits in genetically homogeneous populations, with straightforward generalisations to arbitrary phenotypic spaces with multiple genes.

{\bf Acknowledgments.} AMW is supported by a Marie Curie Career Integration Grant and ERC Starting Grant. 

\appendix

\section{Solution for linear selection}\label{appA}
To calculate the steady state distribution for a linear selection pressure $s(n)=s n$, we define the generating function for the probability distribution, as $G(z)=\sum_n z^n p_n$. In generating function space and steady state, assuming we know $\av{s(n)}=s \av{n}$, Eq.~\ref{master} becomes:

\beq
(bz-\tilde{b})G-(\tilde{d}z-d) \frac{d G}{d z}=0,
\label{gen1}
\eeq

where $\tilde{b}=b+s\av{n}$ and $\tilde{d}=d-s$. Eq.~\ref{gen1} can be solved by direct integration to give
\beq
G(z)=e^{\frac{b}{\tilde{d}}(z-1)} \left( \frac{1-\delta z}{1-\delta} \right)^{\tilde{\beta}},
\eeq
where $\delta=\frac{\tilde{d}}{ d}$ and $\tilde{\beta}= s\frac{{b-d\av{n}+s\av{n}}}{(d-s)^2}$. Note that this expression for the generating function self-consistently satisfies $G'(1)=\<n\>$, so that $\<n\>$ is not constrained by the condition of stationarity. We thus have a family of solutions, parametrized by $\<n\>$ or equivalently by $\tilde \beta$. An especially simple solution is given by the condition $\tilde \beta=0$, which yields the generating function of a Poisson distribution:
\beq
G(z)=e^{\frac{b}{d-s}(z-1)}.
\eeq
This solution is the one we obtain by numerical integration of Eq.~\ref{master} at steady-state.

The Fokker-Planck equation for the continuous case (Eq.\ref{eq:FP}) is solved at steady state by going to Fourier space:
\beq
\tilde P(p)=\int dx e^{ipx}P(x).
\eeq
Eq.\ref{eq:FP} then becomes at steady state (we set $x_0=0$ with no loss of generality):
\beq
(p+is)\partial_p \tilde P+(Dp^2+s\bar x)\tilde P=0,
\eeq
where $\bar x=\int dx\, xP(x)$.
The solution to this equation is:
\beq
\tilde P(p)=e^{-\frac{1}{2}\frac{Dp^2}{k}-\frac{iDps}{k^2}} {\left(1-\frac{pk}{s}\right)}^{s(Ds/k^2-\bar x)/k}.
\eeq
As in the discrete case, the only stable solution corresponds to $\bar x=Ds/k^2$. In this case, the exponent in the second term cancels, and we obtain the Fourier transform of a Gaussian distribution of mean $Ds/k^2$ and variance $D/k$.

\section{Solution with self-regulation}\label{appB}
Here we give some details for the calculations of self-regulation
(Sec.~\ref{regulation}). The birth rate is assumed to depend on $n$:
$b+b_1n$. Using the generating function technique, we get the steady-state solution in absence of selection:
\beq\label{reg}
\frac{G'(z)}{G(z)}=\frac{b}{d-b_1z},
\eeq
from which we infer:
\beq
P(n)={\left(1-\frac{b_1}{d}\right)}^{b/b_1} \frac{1}{n!} {\left(\frac{b}{d}\right)}^n \prod_{i=0}^n \left(1+i\frac{b_1}{b}\right).
\eeq
With selection the equation for the generating function $G(z)$ reads:
\beq
\frac{G'(z)}{G(z)}=\frac{b+s\av{n}-bz}{(d-b_1z)(1-z)+sz}.
\eeq
The right-hand side has two poles:
\beq
z_{\pm}=\frac{b_1+d-s\pm\sqrt{(b_1+d-s)^2-4b_1d}}{2b_1}
\eeq
By analogy with the unregulated case, we make the hypothesis that the only stable solution is such that
the pole at $z_-$ disappears. This is satisfied if:
\beq
\av{n}=\frac{2b}{d-s+b_1+\sqrt{(b_1+d-s)^2-4b_1d}}.
\eeq
Then we simply have:
\beq
\frac{G'(z)}{G(z)}=\frac{b}{b_1(z_+-z)},
\eeq
so we get the same form as Eq.~\ref{reg}, after replacing $d$ by
\beq
\hat d=b_1z_+=\frac{1}{2}\left(d-s+b_1+\sqrt{(d+b_1-s)^2-4b_1d}\right).
\eeq
We checked numerically that our assumption about the cancelation of the $z_-$ pole in $G'/G$ was correct.

\section{Solution for cliff selection}\label{appC}

This appendix contains details of the calculations in the model of
cliff selection. We start with the discrete case, for which the evolution
equation reads:
\beq
\partial_t p_n = b (p_{n-1}-p_n)+d((n+1) p_{n+1}-np_n+\beta )p_n,
\eeq
for $n\geq n_c$ and $p_n=0$ for $n<n_c$.
$\beta=n_cp_{n_c}$. The last term comes from the normalisation
condition and compensates the loss of cells off the cliff, which
happens with rate $d\beta$.
The generating function can be calculated as a function of $\beta$ at
steady state:
\beq
G_{\beta}(z)=e^{\alpha z}(1-z)^{\beta}\int_0^z dy \beta y^{n_c-1}e^{-\alpha y}(1-y)^{-\beta-1},
\eeq
where $\alpha=b/d$. Note that the form above automatically satisfies
$G(z)\sim p_{n_c}z^{n_c}$ as $z\to 0$, and $G(1)=\sum_n
p_n=1$. Therefore $\beta$ is unconstrained and entirely determines the
solution. Guided by numerical simulation, we hypothesize that the only
stable solution corresponds to the highest possible $\beta$ that does
not entail $p_n<0$ for some $n$.
An analogous analytical solution exists for the threshold model, with an
additional continuity condition between the two intervals $(0,n_c-1)$
and $(n_c,+\infty)$.

In the continuous case, the Fokker-Planck equation reads:
\beq
\partial_t P = k\partial_x\left(xP\right) + D\partial_x^2 P +D\beta P,
\eeq
with $\beta=\partial_x P|_{x=x_c}$.
The last term corresponds to the flux of cells crossing the threshold.
The formal solution reads:
\beq
P_{\beta}(x)=xe^{-y(x)}\frac{u(y_c)m(y(x))-m(y_c)u(y(x))}{u(y_c)N_m-myx_c)N_u},
\eeq
where $y(x)=kx^2/2D$, $y_c=kx_c^2/2D$, $N_a=\int_{x_c}^{+\infty}dx\, xe^{-y(x)}u(y(x))$, with $a=u$ or $m$. The functions $m(x)$ and $u(x)$ are defined as: $m(x)=M((k-\beta D)/2k,3/2,x)$ and $m(x)=U((k-\beta D)/2k,3/2,x)$, where $M$ and $U$ are the confluent hypergeometric functions of the first and second kind, respectively.

\bigskip

\section{Optimal switching path}\label{appD}
Here we detail the calculation of the optimal reaction path of a stochastic process under selective pressure. We assume that all cells are equilibrated in one metastable state, and we consider the probability of rare paths out of this state.
The probability of a path is given by the usual expression for the action, multiplied by a term reflecting the historical fitness of the cell relative to the rest of the population \cite{KusselLeibler2010}:
\beq
\begin{split}
P(\{x(t)\}\sim &\exp\left[\int_{t_{\rm initial}}^{t_{\rm final}} dt \left( -\frac{[\frac{dx}{dt} - f(x)]^2}{4D(x)} +s(x) - \<s\>\right)\right]\\
\sim &\exp\left(-\int_{t_{\rm initial}}^{t_{\rm final}} dt\, \mathcal{L}\right).
\end{split}
\eeq
The Lagrangian $\mathcal{L}$ can be rewritten as:
\beq
\mathcal{L}=\frac{(\frac{dx}{dt}-g(x))^2}{4D(x)} + \frac{dx}{dt}\frac{g(x)-f(x)}{2D(x)},
\eeq
with $g(x)=\pm \sqrt{f(x)^2-4D(x)(s(x)-\<s\>)}$. Note the second term in the integrand does not depend on the particular path taken, and that the first term can be made arbitrarily small by setting $dx/dt=g(x)$ and by choosing the sign of $g$ appropriately \cite{bialek,sneppen2}. We are considering rare paths, which move against the drift, {\em e.g.} $dx/dt>0$ and $f(x)<0$. Then the action of the optimal path reads:
\beq
\mathcal{A}=-\int_{x_{\rm initial}}^{x_{\rm final}} dx\, \frac{|f(x)|+\sqrt{f(x)^2-4D(x)(s(x)-\av{s})}}{2D(x)}.
\eeq

\bibliographystyle{biochem}
\bibliography{genesel1}

\end{document}